\documentclass[reprint, amsmath,amssymb, aps, pra,superscriptaddress,email]{revtex4-2}

\usepackage{graphicx}
\usepackage{comment}
\usepackage{amsmath}
\usepackage{color}
\usepackage{hyperref}
\usepackage[T1]{fontenc}
\usepackage[utf8]{inputenc}

\usepackage{soul}

\begin{document}

\preprint{APS/123-QED}

\title{Collective atom-cavity coupling and non-linear dynamics with atoms with multilevel ground states}

\author{Elmer Suarez}
\affiliation{Center for Quantum Science and Physikalisches Institut, Eberhard-Karls Universität Tübingen, Auf der Morgenstelle 14, 72076 Tübingen, Germany}
\author{Federico Carollo}
\affiliation{Institut f\"ur Theoretische Physik, Universit\"at Tübingen, Auf der Morgenstelle 14, 72076 T\"ubingen, Germany}
\author{Igor Lesanovsky}
\affiliation{Institut f\"ur Theoretische Physik, Universit\"at Tübingen, Auf der Morgenstelle 14, 72076 T\"ubingen, Germany}
\affiliation{School of Physics and Astronomy and Centre for the Mathematics and Theoretical Physics of Quantum Non-Equilibrium Systems, The University of Nottingham, Nottingham, NG7 2RD, United Kingdom}
\author{Beatriz Olmos}
\affiliation{Institut f\"ur Theoretische Physik, Universit\"at Tübingen, Auf der Morgenstelle 14, 72076 T\"ubingen, Germany}
\affiliation{School of Physics and Astronomy and Centre for the Mathematics and Theoretical Physics of Quantum Non-Equilibrium Systems, The University of Nottingham, Nottingham, NG7 2RD, United Kingdom}
\author{Philippe W. Courteille}
\affiliation{ Instituto de Física de São Carlos, Centro de pesquisa em óptica é fotônica, Universidade de São Paulo, Brazil }
\author{Sebastian Slama}
\affiliation{Center for Quantum Science and Physikalisches Institut, Eberhard-Karls Universität Tübingen, Auf der Morgenstelle 14, 72076 Tübingen, Germany}
\email{sebastian.slama@uni-tuebingen.de}

\date{\today}

\begin{abstract} 
We investigate experimentally and theoretically the collective coupling between atoms with multilevel ground state manifolds and an optical cavity mode. In our setup the cavity field optically pumps populations among the ground states. The ensuing dynamics can be conveniently described by means of an effective dynamical atom-cavity coupling strength 
that depends on the occupation of the individual states and their coupling strengths with the cavity mode. 
This leads to a dynamical backaction of the atomic populations on the atom-cavity coupling strength which results in a non-exponential relaxation dynamics. We experimentally observe this effect with laser-cooled $^{87}$Rb atoms, for which we monitor the collective normal-mode splitting in real time.
Our results show that the multilevel structure of electronic ground states can significantly alter the relaxation behavior in atom-cavity settings as compared to ensembles of two-level atoms.
\end{abstract}

\maketitle
\section{Introduction}
\begin{figure}[h]
\includegraphics{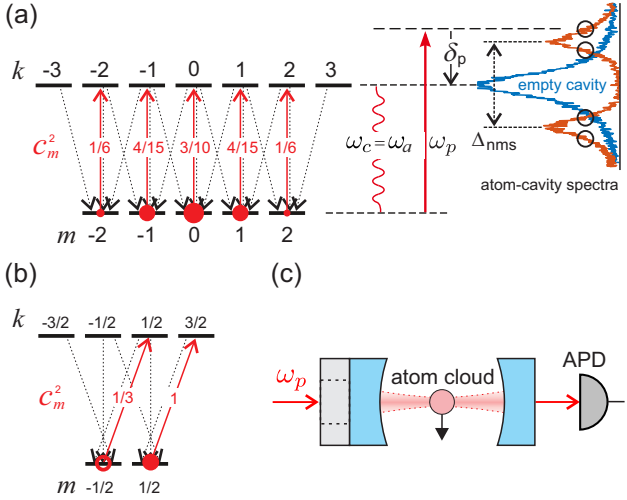}
\caption{\label{fig1}\textbf{Multilevel ground state systems coupled to a single mode cavity.} (a) In our experiment, the $5S_{1/2},F=2$ ground states are coupled to the $5P_{3/2},F'=3$ states by a linearly polarized cavity field driving $\pi$-transitions with Clebsch-Gordan coefficients $c_m$. The measured atom-cavity spectrum indicates the detunings $\delta_p$ used in the experiment (black open circles). (b) The model system describes a simpler situation with two ground states. Here, the cavity field drives $\sigma^+$-transitions. The red full dots in (a) and (b) indicate the steady-state occupations, and the red circle in (b) the initial occupation. (c) The sketch of the experiment shows a free falling cloud of cold $^{87}$Rb atoms inside the cavity, pumped with probe laser frequency $\omega_p$. The transmission is detected on an avalanche photo diode (APD).}
\end{figure}

Many cavity QED experiments focus on implementing ideal toy models by coupling effective two-level atoms to single light modes. However, only few setups have taken the multilevel structure of ground and excited states of realistic atoms into account \cite{Birnbaum2006,Samutpraphoot2020}. In general, the coupling strength depends on the atomic transition coupled to the cavity field. Previous work has exploited this effect to generate spin squeezing of hyperfine ground state levels due to AC Stark shifts that depend on the collective spin \cite{Leroux2010, Hosten2016}. This principle has also been used in the context of the Dicke phase transition including spin-degrees of freedom \cite{Zhiqiang17}. Recent work has observed the formation of spin textures \cite{landini2018}, spin-dependent interactions \cite{Norcia2018,davis2019,guo2019}, and ground-state bistability \cite{Domokos2022}. Despite this progress, collective coupling of atoms with magnetic sublevels in the ground and excited state, illustrated in Fig.~\ref{fig1}, is widely unexplored. Here, the coupling strength depends on the individual transitions via their Clebsch-Gordan coefficients, and complex relaxation dynamics to the steady state can arise when sublevel-changing processes (pumping) influence the collective coupling strength \cite{Arnold2011}. Such complex dynamics has been proposed for dissipative many-body quantum systems with interactions \cite{Poletti2012,Cai2013,Chiacchio2018}. Moreover, superradiance in this kind of system can lead to the population of long-lived dark states, as has been recently proposed in \cite{Orioli2022}.

In this work we investigate the dynamics of an ensemble of atoms with degenerate ground state manifold when coupled to a single-mode cavity (Fig.~\ref{fig1}). We derive how the collective coupling strength depends on the population of the individual sublevels and experimentally detect the collective atom-cavity coupling strength in real-time. Moreover, we show that the backaction of the coupling strength on the intracavity light field can lead to non-linear dynamics within the manifold of atomic ground states. This non-linear behavior is exclusively caused by the multilevel structure and is absent in the case of two-level atoms. In particular, it does not require strong pumping nor mechanical backaction, which are well-studied mechanisms that lead to nonlinear dynamics in atom-cavity systems \cite{Orozco1997, Ritsch2013}.

\section{Multilevel collective coupling}
The hallmark of collective atom-cavity coupling is the observation of a collective normal mode splitting in the light transmitted through the cavity. A number $N$ of atomic two-level systems resonantly coupled to the cavity with coupling constant $g_0$ splits the cavity transmission into two peaks with separation \cite{Gripp1997}
\begin{equation}
\label{eq:nms}
 \Delta_\mathrm{nms}=2g_0\sqrt{N}. 
\end{equation}
This expression is valid in the weak-field and collective strong coupling regime, where $g_0^2N\gg\Gamma\kappa$, with $\Gamma$ and $\kappa$ being the atomic excited state and cavity field decay rate, respectively. Atom number variations can thus influence the transmission through the cavity. We have recently used this effect to detect Rydberg excitation dynamics in the cavity in real-time \cite{Suarez2022}. Note, that the atom number $N$ is in general an effective atom number that takes the intensity profile of the mode function and the positions of the atoms into account. This can give rise to optomechanical effects, such as optomechanical cooling, optical instabilities, and phase transitions where atoms are arranged in complex structures \cite{Ritsch2013,Labeyrie2014,Ackemann2022}. In this work, we consider only weak cavity fields with negligible mechanical backaction and thus treat $N$ as a constant parameter. 

The collective character of the normal mode splitting is based on the excitation of a coherent Dicke state at the lower end of the Dicke ladder, where a number $n_\mathrm{ph}\ll N$ of cavity photons is equally shared among a much bigger number $N$ of atoms \cite{Garraway2011,Kirton2017,Kirton2019}. Here, we assume that initially the atomic population is spread over a number $n_\mathrm{gs}$ of independent ground states. Each ground state is coupled by the cavity field to an individual excited state, as shown in Fig.~\ref{fig1}(a) for the Rb $5S_{1/2},F=2$ to $5P_{3/2},F'=3$ transition. The cavity field is pumped by an external laser field with rate $\eta$, and frequency $\omega_p$, detuned from the atom transition frequency $\omega_a$ and cavity resonance frequency $\omega_c$ by $\Delta_{a,c}=\omega_p-\omega_{a,c}$ respectively. In our experiment, the probe laser detuning is $2\pi\delta_p=\Delta_a=\Delta_c$ and the linearly polarized cavity field couples only states with $\Delta m=0$, but similar models can be derived for any transition and polarization state. An example with circularly polarized light is shown in Fig.~\ref{fig1}(b), where only $\Delta m=+1$ transitions are driven. For each ground state $\left|g,m\right>$, a single coupling constant $g_{m}$ can be defined by     
\begin{equation}
\label{eq:gj}
g_m=\mu_m\sqrt{\frac{\omega_a}{\hbar\varepsilon_0V_\mathrm{mode}}},
\end{equation}
with cavity mode volume $V_\mathrm{mode}$. The dipole matrix element $\mu_m$ of the corresponding transition can be expressed as $\mu_m=\mu_\mathrm{red}c_m$, with reduced dipole matrix element $\mu_\mathrm{red}$ and Clebsch-Gordan coefficient $c_m$ \cite{Steck2021}. Thus the individual coupling constants are proportional to the Clebsch-Gordan coefficients of the corresponding transition, i.e.
\begin{equation}
\label{eq:gj2}
g_m=g_0c_m,
\end{equation}
with $g_0$ defined by \eqref{eq:gj} and \eqref{eq:gj2}.

In order to identify how the normal mode splitting behaves in the multilevel case, the usual atom-cavity Hamilton operator \cite{Gripp1997} is extended to
\begin{equation}
\label{eq:Hamilton1}
\hat{H}=\hat{H}_0 +\hbar\sum_{j=1}^{N}\sum_{m}g_m (i\hat{a}^\dagger \hat{\sigma}_{jm} + h.c.),
\end{equation}
where $m$ runs over the magnetic quantum number of each ground state, $\hat{a}^\dagger$ and $\hat{a}$ are the bosonic raising and lowering operators of the cavity field, and $\hat{\sigma}_{jm}=\left|g,m\right>_j\left<e,m'\right|$ is the dipole operator for the transition driven by the cavity only, i.e. $m'=m$ and $m'=m\pm1$ for a linearly and circularly polarized cavity field, respectively. The first term of the Hamiltonian, 
\begin{equation}
\hat{H}_0=-\hbar\Delta_a\sum_{j,m}\hat{\sigma}_{jm}^\dag\hat{\sigma}_{jm}-\hbar\Delta_c\hat{a}^\dagger\hat{a}+i\hbar\eta(\hat{a}^\dagger-\hat{a})    
\end{equation}
is the uncoupled Hamiltonian of the atoms and the cavity field. The second term of (\ref{eq:Hamilton1}) describes the coherent coupling of atom $j$ in state $m$ and the cavity field. Here, having neglected the position dependence $e^{-i\vec k\cdot\vec r_j}$ of the atoms, we assume that all atoms are placed at the positions of maximum coupling strength. Other atomic density distributions can be considered in the weak-field limit by introducing an effective atom number \cite{Ritsch2013}.

Including the atomic and cavity field decay, the dynamics of the system is described by the master equation
\begin{eqnarray}
\label{eq:masterequation}
\dot\rho&=&-\frac{i}{\hbar}[\hat H,\rho]\nonumber\\
&&+\Gamma\sum_{j,m}\sum_{k=m-1}^{m+1}\left(\hat{L}_{mk}^j\rho \hat{L}^{j\dagger}_{mk}-\frac{1}{2}\left\{\hat{L}^{j\dagger}_{mk}{\hat{L}^j_{mk}},\rho\right\}\right)\nonumber\\
&&+2\kappa\left(\hat{a}\rho\hat{a}^{\dagger}-\frac{1}{2}\left\{\hat{a}^{\dagger}\hat{a},\rho\right\}\right).
\end{eqnarray}
Note, that unlike in the atom-cavity coupling where each ground state is coupled to a single excited state depending on the chosen cavity field polarization, the spontaneous decay couples all allowed dipole transitions. This is well justified when the spontaneous emission is not influenced by the presence of the cavity whose Purcell factor is small \cite{Heinzen1987}. Hence, the spontaneous decay jump operators for the atom $j$ reads
\begin{align}
\label{eq:jumpops1}
\hat{L}_{mk}^j=\sqrt{\beta_m^{k}}\left|g,m\right>_j\left<e,k\right|,
\end{align}
where the factors $\beta_m^k$ describe the branching ratio of the decay probability from the excited state $\left|e,k\right>$ to the ground state $\left|g,m\right>$. The equations of motion for the expectation values of the cavity and atomic operators can be derived from Eqns. \eqref{eq:Hamilton1}\---\eqref{eq:jumpops1}. Neglecting atom-photon correlations, i.e. making a meanfield approximation, the equations read
\begin{align}
\label{eq:pde_a}
&\dot{a}=-(\kappa-i\Delta_c)a+\sum_{j,m}g_m\sigma_{jm}+\eta\\
\label{eq:pde_sigma}
&\dot{\sigma}_{jm}=-\left(\frac{\Gamma}{2}-i\Delta_a\right)\sigma_{jm}+g_m\sigma_{jm}^z a\\
\label{eq:pde_P}
&\dot{P}_{jm}=-\Gamma\rho_{jm'}^{ee}+\Gamma\sum_{k=m-1}^{m+1}\beta_m^k\rho_{jk}^{ee}\\
\label{eq:pde_rho}
&\dot{\rho}_{jm'}^{ee}=-\Gamma\rho_{jm'}^{ee}-g_{m}\left(a\sigma_{jm}^\dag+a^*\sigma_{jm}\right).
\end{align}
Here, we have introduced the quantity $P_{jm}=\rho_{jm}^{gg}+\rho_{jm'}^{ee}$ as the probability that atom $j$ participates in transition $m\to m'$ driven by the cavity, with $\rho_{jm}^{gg}$ and $\rho_{jm'}^{ee}$ being the probability of the $j$-th atom being in the ground state $\left|g,m\right>_j$ and excited state $\left|e,m'\right>_j$, respectively. 
Its values are changed by spontaneous decay with rates $\Gamma\beta_m^k$. The inversion of atom $j$ within transition $m$ is given by $\sigma_{jm}^z=\rho_{jm'}^{ee}-\rho_{jm}^{gg}=2\rho_{jm'}^{ee}-P_{jm}$. 

\begin{figure}[t]
\includegraphics{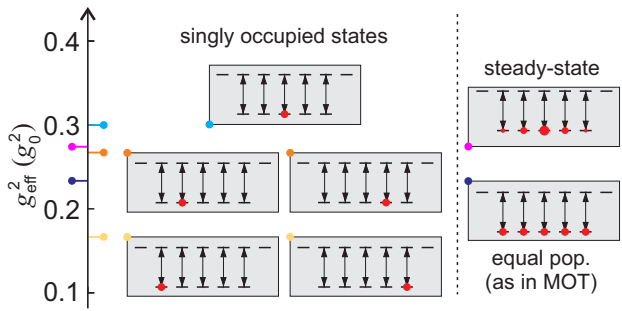}
\caption{\label{fig2}\textbf{Effective coupling strength.} $g_\mathrm{eff}^2$ in units of the coupling constant for various ground-state populations with cavity-driven $\pi$-transitions. The effective coupling strength of the steady-state is larger than that of the state with equal populations.}
\end{figure}

We assume that all atoms participating in transition $m\to m'$ follow the same internal dynamics, such that $\sigma_{jm}=\sigma_{j'm}\equiv\sigma_m$, $P_{jm}=P_{j'm}\equiv P_m$, and $\rho_{jm'}^{ee}=\rho_{j'm'}^{ee}\equiv\rho_{m'}^{ee}$. 
Thus, Eqns. \eqref{eq:pde_a}\---\eqref{eq:pde_rho} form a set of only $3n_\mathrm{gs}+1$ differential equations, compared to the full set of $3Nn_\mathrm{gs}+1$ equations. The steady-state solution of the intracavity power obtained from Eqns. \eqref{eq:pde_a}\---\eqref{eq:pde_rho} within the weak-field approximation (assuming that in the steady-state $\rho^{ee}_{m'}=0$ for all $m'$) is given by
\begin{equation}
\label{eq:a_2}
    \left|a\right|^2=\frac{\left(\frac{\Gamma^2}{4}+\Delta_a^2\right)\eta^2}{N^2g_\mathrm{eff}^4\!+\!Ng_\mathrm{eff}^2\left(\Gamma\kappa\!-\!2\Delta_a\Delta_c\right)\!+\!\left(\frac{\Gamma^2}{4}\!+\!\Delta_a^2\right)\!\left(\kappa^2\!+\!\Delta_c^2\right)},
\end{equation}
with the effective coupling strength
\begin{equation}
\label{eq:g_eff}
    g_\mathrm{eff}^2=g_0^2\sum_{m}{{c_m^2P_m}}.
\end{equation}
Note, that $g_\mathrm{eff}$, like $P_m$, is in general a time-dependent quantity. As the individual Clebsch-Gordan coefficients are typically different for different transitions, the effective coupling strength depends on the population of the individual ground state levels. So does the size of the normal mode splitting which is derived from Eq. \eqref{eq:a_2} as
\begin{equation}
\label{eq:nms_eff}
\Delta_\mathrm{nms}=2g_\mathrm{eff}\sqrt{N},
\end{equation}
valid under the same approximations as in Eq. \eqref{eq:nms}. The fact that the normal mode splitting is proportional to $\sqrt{N}$ means that all $N$ atoms contribute to a collective Dicke state. Indeed, the indistinguishable nature of the atoms exchanging photons with the cavity field implies that the atomic ensemble forms a Dicke state. However, in our case, this state is further partitioned following the statistical distribution over the ground state manifold, Eq. \eqref{eq:g_eff}. The effective coupling strength is plotted in Fig. \ref{fig2} for various populations including the situation realized in the experiment, where all levels are initially equally populated. We will see in the following section that the effective coupling strength and the normal mode splitting dynamically change when pumping between the ground state levels occurs. 

\section{Experiment}
In order to detect the dependence of the normal mode splitting on the distribution among different sub-levels, a cold cloud of $^{87}$Rb atoms is prepared in a magneto-optic trap (MOT) and positioned in the mode volume of a near-confocal multi-mode cavity with round-trip length $l=10~\mathrm{cm}$ and finesse ${\cal F}=224$, such that the full-width at half-maximum is $\nu_\mathrm{fwhm}=13.4$~MHz [see Fig.\ref{fig1}(c)]. The single atom-cavity coupling is $g_0=2\pi\times210$~kHz, further experimental details are given in \cite{Suarez2022}. Within the MOT, all sublevels of the $5S_{1/2}, F=2$ ground state are assumed to be equally populated. The atoms are then released from the MOT by switching off the MOT lasers and accelerated due to gravity. Simultaneously, the magnetic fields operating the MOT are switched off and, instead, are tuned to compensate the Earth magnetic field. Thus, all Zeeman sublevels are degenerate. The cavity length is stabilized by a far-detuned lock laser such that the cavity is resonant with the atomic transition from the ground to the  $5P_{3/2}, F'=3$ excited state. In order to detect the normal mode splitting, a weak, linearly polarized probe light field is coupled into the cavity and detected in transmission on an avalanche photodiode. The probe frequency is detuned with respect to the transition by $\delta_p$ to the side of the fringe of one of the normal modes, see Fig.~\ref{fig1}(a). That way, the transmission detects changes of the normal mode splitting.

\begin{figure}[t]
\includegraphics[scale=1]{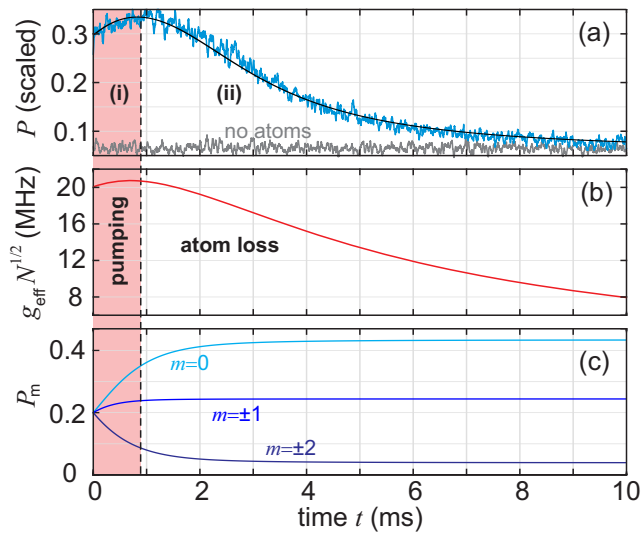}
\caption{\label{fig3}\textbf{Measured dynamics.} (a) Measured and simulated cavity transmission for probe detuning  $\delta_p=+24$~MHz, scaled to the empty cavity maximum of $P_\mathrm{cav}=2.4$~nW and an atom number of $N=11,200$. Two time intervals can be identified, where the signal is mainly dominated by (i) redistribution of atoms among the sub-levels by pumping and (ii) loss of atoms from the cavity. (b) Dynamics of the collectively enhanced effective coupling strength corresponding to the simulation in (a). (c) Time evolution of the populations $P_m$ of the ground state levels, corresponding to the simulation in (a). }
\end{figure}
\begin{figure}
\includegraphics[scale=0.95]{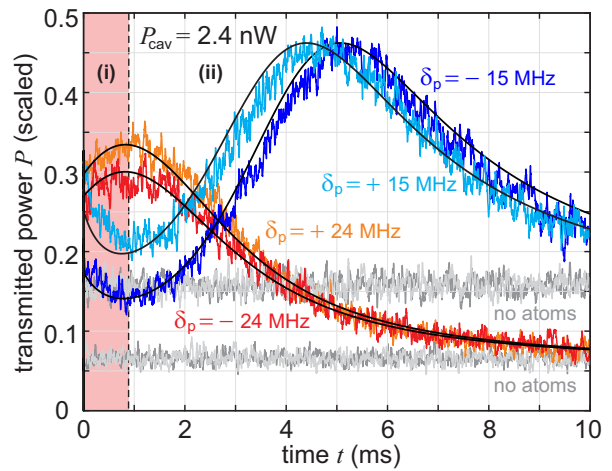}
\caption{\label{fig4}\textbf{Detuning-dependent transmitted power.} Measured and simulated cavity transmission as explained in Fig.~\ref{fig3}(a) for different values of $\delta_p$. Fitted atom numbers are $N_\mathrm{-24MHz}=10,600$, $N_\mathrm{-15MHz}=9,200$, $N_\mathrm{+15MHz}=10,500$, and $N_\mathrm{+24MHz}=11,200$ for the corresponding detunings $\delta_p$. The fitted intracavity power $P_\mathrm{cav}=2.4$~nW is equal for all curves.}
\end{figure}

The measured transmitted power $P$ is plotted in Fig.~\ref{fig3}(a) for a detuning of $\delta_p=+24$~MHz, at which the probe laser is set to the outer slope of the normal mode splitting. We observe that it first increases during time interval (i) and then decreases during (ii). These two regimes can be better understood studying the dynamics obtained by solving the Eqns. (\ref{eq:pde_a})-(\ref{eq:pde_rho}). In Fig.~\ref{fig3}(a) we find that the transmitted power $P\propto|a|^2$ coincides remarkably well with the measured one. Note, that the intracavity power (empty cavity, on resonance) is fitted to $P_\mathrm{cav}=2.4$~nW, agreeing very well with the value of $P_\mathrm{cav}=2.7$~nW that we determine from the measured power transmitted through the cavity, with $1.5\%$ transmission of the outcoupling mirror of the cavity. Simulating the dynamics of the effective coupling strength $g_\mathrm{eff}^2$ and the population of the ground state levels $P_m$ [Figs.~\ref{fig3}(b) and (c), respectively], we can see that the signal is dominated in the regime (i) by the change in the effective coupling strength due to a redistribution of the populations. Pumping with linearly polarized light that drives $\pi$-transitions, leads to a higher occupation of sub-levels with small $m$ quantum numbers, as shown in Fig.~\ref{fig3}(c). These levels are coupled with larger Clebsch-Gordan coefficients, leading to a collectively enhanced effective coupling strength, which increases by approximately 1~MHz [see Fig.~\ref{fig3}(b)]. Thus, $\Delta_\mathrm{nms}$ increases and the peak of the normal mode splitting is shifted to higher frequencies, getting closer to the probe frequency, by which the transmission increases. For longer times, during time interval (ii), atoms are lost from the cavity mode volume due to free expansion of the cloud. Since $\Delta_\mathrm{nms}$ is reduced $\propto\sqrt{N}$, this leads to the observed decrease of cavity transmission in (ii).

Similar dynamics can be observed when the probe laser is tuned to one of the other three slopes of the normal mode splitting, as shown in Fig.~\ref{fig4}. The dynamics of the cavity transmission at the inner sides of the normal mode splitting ($\delta_p=\pm15$~MHz), i.e.~the blue measurements shown in Fig.~\ref{fig4}, have an extra feature. There, the initial increase of the $\Delta_\mathrm{nms}$ in region (i) leads to a decrease in the transmitted power. The subsequent decrease of $\Delta_\mathrm{nms}$ during (ii) increases the transmission correspondingly, until a maximum is reached when the peak frequency of the splitting coincides with the probe frequency. After that, the transmission goes back to the value of the empty cavity. All data curves are compared again with simulations (black lines), where the atom number in each curve is fitted separately.

In our simulations the motion of the atoms is described as ballistic expansion under the action of gravity with a fixed temperature of the atoms of $T=75~\mu$K. In the experiment, the motional dynamics is more complicated, as scattering of probe light heats up the atoms in the cavity. We do not simulate this temperature change, as we are only interested in the internal state dynamics and its influence on the collective coupling strength at early times. In order to exclude mechanical action of the standing light wave on the atoms, we calculate the depth of the dipole potential at an antinode to be $U=k_B\times 0.5~\mu$K \cite{Grimm2000}, with a laser detuning of $\delta_p=25$~MHz, an intracavity laser power of $P_\mathrm{cav}=2.4$~nW, and a beam waist of $w_0=80~\mu$m. The potential depth is thus substantially smaller than the temperature of the atom cloud. We have also checked by inspecting simulated trajectories of atoms, that they are only little influenced by the optical dipole potential.

\begin{figure}[t]
\includegraphics[scale=1]{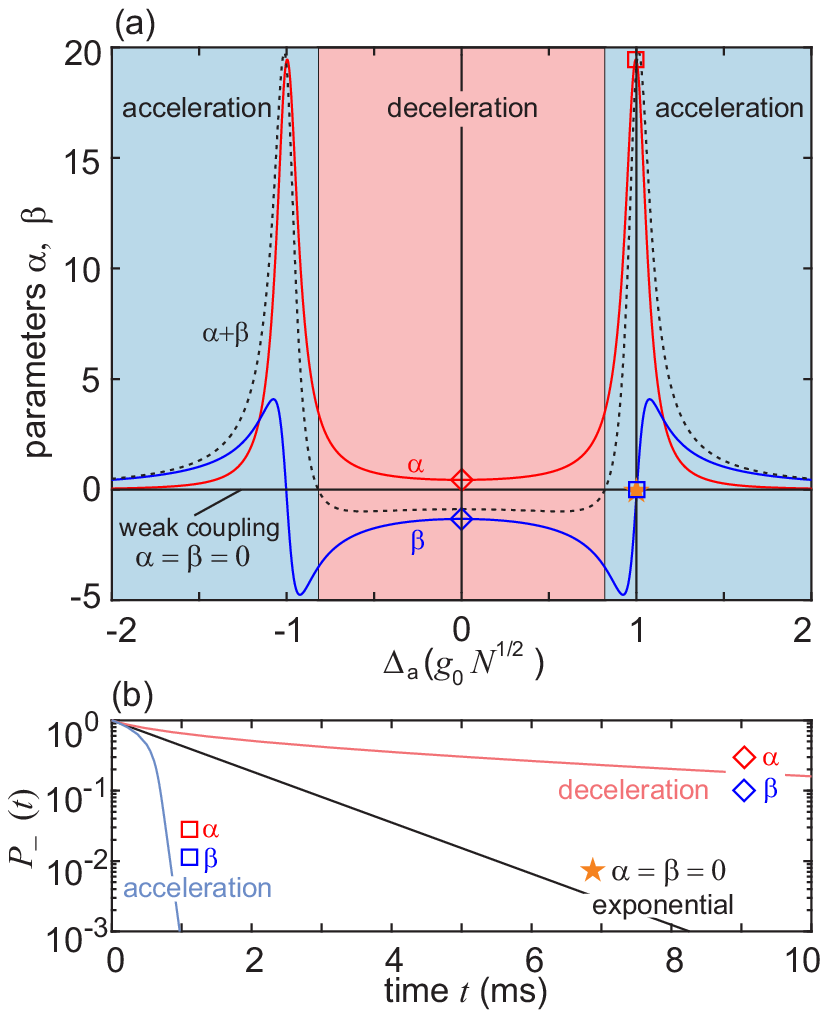}
\caption{\label{fig5}\textbf{Nonlinear dynamics.} (a) Parameters $\alpha$ and $\beta$ appearing in differential equation \eqref{eq:rate_P0}, calculated for strong coupling with $g_0\sqrt{N}/\Gamma=10$ (red and blue solid lines) and weak coupling with $g_0\sqrt{N}/\Gamma=0.01$ (black solid lines; on top of each other). In the simulation we set $\Gamma=\kappa$ and $\Delta_c=\Delta_a$. The black dashed line is the sum $\alpha+\beta$ and determines whether the initial dynamics for $P_{-}=1$ is accelerated (for positive values) or slowed down (for negative values). (b) Simulated time dynamics of $P_{-}(t)$ for weak and strong coupling. In the weakly coupled case (black curve) the decay follows an exponential function. Strong coupling can lead to acceleration (cyan) or deceleration (pink) of this dynamics. The corresponding detunings of $\Delta_a=0$ and $\Delta_a=g_0\sqrt{N}$ have been chosen as indicated by the symbols in (a). The laser power in the different cases has been adjusted such that all curves start with the same slope.}
\end{figure}

\section{Non-linear dynamics}
In this section we focus on the dynamics of the population of the atomic ground states. We will perform a theoretical investigation that shows how the multilevel structure of the ground state manifold is responsible for non-linear effects. To illustrate the underlying mechanism we resort to the simple two-transition model shown in Fig.~\ref{fig1}(b). Initially, all atoms are prepared in the $m=-1/2$ level, and a circularly polarized pumping field pumps all atoms to the $m=+1/2$ level. Feedback is generated by the interaction with the cavity: the strength of the cavity field depends via the collective coupling strength on the exact population of the individual levels and is thus dynamically varying. The non-linearity becomes apparent by deriving rate equations from the equations of motion \eqref{eq:pde_a}\--\eqref{eq:pde_rho} in the weak driving limit. Here, the populations are slowly evolving while the cavity is adjusting instantaneously, and the splitting follows proportionally to $g_\mathrm{eff}$. We separate the timescales by setting $\dot{a}=\dot\sigma_{jm}=\dot\rho^{ee}_{jm}=0$ and include weak pumping by making the approximation $\sigma_{jm}^z(t) \approx -P_{jm}(t)$ in Eq. \eqref{eq:pde_sigma}. Assuming again that all atoms follow the same internal dynamics, the rate equations for the population of state $m=-1/2$, $P_-$, for this two-transition model are then given by
\begin{equation}
\label{eq:rate_P0}
\dot P_-=-\Gamma_\mathrm{eff} f(P_-) P_-,
\end{equation}
with the effective decay rate
\begin{eqnarray}
\Gamma_\mathrm{eff}=&\frac{\eta^2}{g_0^2N^2}\frac{c_{-}^2}{c_{+}^2}\frac{\Gamma}{\left(c_{+}^2+u\right)+w}.
\end{eqnarray}
The effective decay rate can be tuned by the cavity pumping rate $\eta^2$. The equation for $P_+$ follows from $P_-+P_+=1$. The non-linearity of the dynamics is encoded in the function $f(P_{-})$, given by 
\begin{equation}
\label{eq:f(P)}
f(P_{-})=\frac{1}{\alpha P_{-}^2+\beta P_{-} + 1},
\end{equation}
with parameters
\begin{align}
\label{eq:f_P}
\alpha=&\frac{\left(c_{-}^2-c_{+}^2\right)^2}{c_{+}^2\left(c_{+}^2+u\right)+w}\\
\beta=&\frac{2\left(c_{-}^2-c_{+}^2\right)\left[1+\frac{u}{2}\right]}{\left(c_{+}^2+u\right)+w},
\end{align}
where
\begin{align}
\label{eq:u_und_w}
u=&\frac{\Gamma\kappa-2\Delta_a\Delta_c}{g_0^2N},\\
w=&\frac{\left[\left(\frac{\Gamma}{2}\right)^2+\Delta_a^2\right]\left[\kappa^2+\Delta_c^2\right]}{g_0^4N^2}
\end{align}
and $c_\pm$ are the Clebsch-Gordan coefficients for the transitions with $m=\pm1/2$, respectively. The dynamics determined by these rate equations yield virtually the same results as the consideration of the full equations \eqref{eq:pde_a}\--\eqref{eq:pde_rho}.

The solution of Eq. \eqref{eq:rate_P0} in its implicit form $t(P_{-})$, using the initial condition $P_-(t=0)=1$, is
\begin{equation}
\label{eq:P0_solution}
t=-\frac{1}{\Gamma_\mathrm{eff}}\left[\frac{\alpha}{2}P_{-}^2+\beta P_{-}+\ln(P_{-})\right]+\frac{\alpha+2\beta}{2\Gamma_\mathrm{eff}}.
\end{equation}

In the following we analyze a number of limiting cases of the dynamics whose functional dependence is determined by the parameters $\alpha$ and $\beta$. If both parameters are substantially smaller than one, Eq. \eqref{eq:rate_P0} reads
\begin{equation}
\label{eq:rate_P0_linear}
\dot P_{-}=-\Gamma_\mathrm{eff} P_{-},
\end{equation}
and the dynamics follow an exponential decay. As both parameters are proportional to $c_{-}^2-c_{+}^2$, this can be experimentally realized by driving $\pi$-transitions, for which the Clebsch-Gordan coefficients in the two-transition model are equal, i.e. $c_{-}^2=c_{+}^2$. As can be seen in Fig.~\ref{fig5}(a), the condition $|\alpha|,|\beta|\ll1$ can also be reached by choosing large detunings $|\Delta_a|\gg g_0\sqrt{N}$, or --- for arbitrary detuning --- by reducing the coupling strength to values $g_0\sqrt{N}\ll\Gamma$ via the atom number in order to be in the weak coupling limit. This is equivalent to the case of pumping the atoms in free space with no backaction on the cavity field strength. The exponential decay of $P_-$ in this weak coupling regime can be observed in Fig.~\ref{fig5}(b).

The dynamics of $P_{-}$ in the strongly coupled cavity regime, on the other hand, are highly non-linear and can be accelerated or decelerated, depending on the sign of $\alpha+\beta$. This is observable in Fig.~\ref{fig5}(b), where we compare the dynamics $P_{-}(t)$ in the weak and strong coupling regime. In order to make a fair comparison, we tune the cavity pumping rate $\eta^2$ such that the initially absorbed light power $\propto \frac{P_\mathrm{cav}(0)}{\Delta_a^2+(\Gamma/2)^2}$ is equal in all cases. Thus, the decay starts with the same slope, and only the influence of $\alpha$ and $\beta$ on the dynamics becomes visible, i.e. acceleration of the dynamics for $|\Delta_a|= g_0\sqrt{N}$ and deceleration for $\Delta_a=0$, where $\alpha+\beta$ is larger and smaller than zero, respectively. The case of $|\Delta_a|=g_0\sqrt{N}$ is particularly interesting, because in this case the parameter $\beta=0$, whereas $\alpha$ reaches a maximum which scales for strong coupling like
\begin{equation}
\label{eq:alpha_scaling}
\alpha=\left(c_-^2-c_+^2\right)^2\frac{g_0^2N}{\left(\frac{\Gamma}{2}+\kappa\right)^2}.
\end{equation}
As can be seen in Fig. \ref{fig5}, the maximum of $\alpha$ reaches values much larger than one, such that differential equation \eqref{eq:rate_P0} can be approximated as
\begin{equation}
\label{eq:rate_P0_nonlin}
\dot P_-=-\frac{\Gamma_\mathrm{eff}}{\alpha} \frac{1}{P_-}=-\frac{1}{2\tau}\, \frac{1}{P_-},
\end{equation}
which is valid as long as $\alpha P_-^2>1$. The solution of Eq. \eqref{eq:rate_P0_nonlin} is not an exponential,
\begin{equation}
\label{eq:P0_solution_sqrt}
P_-(t)=\sqrt{1-\frac{t}{\tau}}.
\end{equation}
The dynamics for sufficiently short times, i.e., for $t$ being much smaller that the characteristic time scale $\tau$, thus follows a square root law. For later times, where $\alpha P_-^2<1$, the third term in the denominator of Eq. \eqref{eq:f(P)} dominates, and the dynamics follows again an exponential decay.

\section{Conclusion}

This paper reveals a non-linearity in a system of atoms inside an optical cavity in the collective strong coupling regime. The non-linearity does not require saturation nor mechanical backaction. Instead, it is only based on the existence of multilevel ground states with unequal Clebsch-Gordan coefficients which results in an effective coupling strength that depends on the occupation of the individual sublevels. Thus, pumping between the sublevels can dynamically change the effective coupling strength and the intracavity field strength. We experimentally observe this dynamics by detecting the cavity transmission. We furthermore show that backaction of the population dynamics on the cavity field strength leads to non-exponential decay that can be accelerated or decelerated compared to when the system is pumped with constant field strength. Based on our findings for weak driving it would be interesting to investigate further the limit of strong driving, including saturation, and see whether atoms with multilevel ground states behave qualitatively different in comparison to two-level atoms.

\begin{acknowledgments}
The project was funded by the Deutsche Forschungsgemeinschaft (DFG, German Research Foundation) - 422447846 and 465199066. It was carried out within Research Unit FOR 5413 "Long-range interacting quantum spin systems out of equilibrium: Experiment, Theory and Mathematics". Ph.W.C.~acknowledges support from FAPESP grant No.~2022/00261-8.
\end{acknowledgments}

\bibliography{references}

\end{document}